\documentclass[conference]{IEEEtran}
\IEEEoverridecommandlockouts

\usepackage{cite}
\usepackage{amsmath,amssymb,amsfonts}
\usepackage{algorithmic}
\usepackage{graphicx}
\usepackage{textcomp}
\usepackage{xcolor}
\usepackage{booktabs}
\usepackage{tabularx}
\usepackage{array}
\usepackage{multirow}
\usepackage{amsmath}
\usepackage{url}
\usepackage{enumitem}
\usepackage{microtype}
\usepackage{balance}
\usepackage[hidelinks]{hyperref}

\newcolumntype{Y}{>{\raggedright\arraybackslash}X}

\usepackage{soul}

\definecolor{DeepGreen}{RGB}{0,128,0}
\sethlcolor{green!30}

\usepackage{tikz}
\usetikzlibrary{arrows.meta,positioning,fit,calc,backgrounds}

\begin{document}

\title{LLM Watermarking as Big Data Provenance:\\ A Deployment-Oriented Systematization
}


\author{
\IEEEauthorblockN{
Huy Phan\textsuperscript{1+},
Kieu Dang\textsuperscript{2+},
Ojaswi Dulal\textsuperscript{2},
Aiham Al Shukairi\textsuperscript{2},
Abby Shine\textsuperscript{2},
Chase Garner\textsuperscript{2},
Phung Lai\textsuperscript{2}\IEEEauthorrefmark{1}
}
\IEEEauthorblockA{
\textsuperscript{1}Truman State University, Kirksville, MO, USA \\
\textsuperscript{2}State University of New York at Albany, Albany, NY, USA \\
\texttt{minhhuyphan2003@gmail.com, vdang@albany.edu, lai@albany.edu} \\
\textsuperscript{+}These authors contributed equally to this work.
}
\thanks{\IEEEauthorrefmark{1} Corresponding author}
}

\maketitle

\begin{abstract}
As large language models (LLMs) generate and transform content at scale, reliable attribution becomes an increasingly important requirement for big data provenance. Although numerous watermarking methods have been proposed, existing evaluations predominantly emphasize sample-level detectability and robustness, leaving deployment requirements such as serving efficiency, population-scale false positives, multi-key attribution, and long-term verification less systematically examined. 
This paper presents a deployment-oriented systematization of LLM watermarking through the five Vs of big data: Volume, Velocity, Variety, Veracity, and Value. We classify representative watermarking methods across four dimensions: insertion point, verification authority, operational state, and adversarial threat model. We further map these methods to four deployment workloads: online generation, streaming detection, transformation pipelines, and ecosystem governance. Our synthesis identifies recurring limitations, including dependencies on generation-time access, additional inference overhead, sensitivity to downstream transformations, and incomplete reporting of calibration and lifecycle compatibility. Based on these findings, we propose workload-specific evaluation requirements covering statistical reliability, throughput, utility, robustness, and governance. This framework connects watermark design choices to practical deployment constraints and provides guidance for selecting and evaluating provenance mechanisms in large-scale data ecosystems.

\end{abstract}


\section{Introduction}
Large language models (LLMs) are increasingly integrated into large-scale data systems that generate, process, store, and redistribute text at high volume and velocity \cite{OpenAI,gemini,zhou2025survey,fernandez2023large}. Their outputs can propagate across heterogeneous environments, including document repositories, social media streams, search indexes, retrieval-augmented generation (RAG) corpora, training datasets, and automated decision pipelines \cite{dai2024neural,shumailov2024ai}. As synthetic content moves through these systems, it may be copied, summarized, translated, or paraphrased, making its origin difficult to verify. This creates a fundamental big data provenance problem: \textit{How can systems reliably track the source and integrity of machine-generated content across large and dynamic data pipelines?} Reliable provenance is important for accountability, intellectual property (IP) protection, data quality, and trustworthy analytics. Providers may need to determine whether content originated from a particular model, whether protected outputs were used in downstream training, or whether generated content was substantially transformed. LLM watermarking has therefore emerged as a promising mechanism for embedding machine-detectable signals into  the resulting text during or after generation to support scalable provenance and attribution across big data ecosystems~\cite{kirchenbauer2023watermark,kuditipudi2023robust,christ2024undetectable,dathathri2024scalable}.

Most LLM watermarking studies \cite{kirchenbauer2023watermark,giboulot2024watermax,molenda2024waterjudge,yoo2024advancing,liu2023unforgeable} focus on algorithmic design, comparing approaches such as token partitioning, distribution-preserving sampling, semantic encoding, lexical substitution, and cryptographic verification. While these distinctions are important, they do not fully address the challenges of deploying watermarking in big data systems. A model service may generate millions of outputs each day\cite{jaiswal2025serving}, while downstream platforms must detect watermarks in heterogeneous, partially modified content originating from different models, domains, and languages. Data repositories may further combine watermarked text with other sources, and external auditors may require statistically reliable evidence without access to a provider's secret keys. Under these conditions, the key question is not simply whether a watermark can be detected on an individual sample, but whether it remains accurate, robust, scalable, and cost-effective throughout the entire lifecycle of large-scale data.

This systems perspective aligns naturally with the classic 5V characterization of big data. \textit{Volume} increases the impact of even small false-positive rates (FPR) when watermark detectors operate over millions of documents. \textit{Velocity} makes embedding and detection overhead important for throughput and latency. \textit{Variety} requires watermarking methods to remain effective across languages, domains, code, mixed human–AI content, and potentially multimodal pipelines. \textit{Veracity} requires well-calibrated detection under distribution shifts and adversarial transformations. Finally, \textit{Value} reflects whether the provenance benefits justify losses in output quality, computational cost, integration complexity, and governance burden. Existing surveys and evaluations provide useful taxonomies and robustness analyses~\cite{liu2024survey,lalai2025intentions,piet2025markmywords}, but the relationship between watermark design choices and these big data operational requirements remains insufficiently unified in existing literature.

To address this gap, we view LLM watermarking as big data provenance infrastructure and organize the literature around the requirements of large-scale data systems rather than watermark algorithms alone. Our study is a deployment-oriented systematization that connects existing watermark designs, reported evaluations, and known limitations to practical data-management requirements. Our contributions are:
\begin{itemize}[leftmargin=*,nosep]
\item \textbf{A 5V-based provenance formulation.} We map \textit{Volume, Velocity, Variety, Veracity, and Value} to concrete watermarking requirements, including population-scale false-positive control, processing latency, heterogeneous data coverage, robustness under transformation, and deployment cost.
\item \textbf{A deployment-oriented taxonomy.} We organize watermarking methods along four complementary dimensions: where the watermark is embedded, who can verify it, what operational state or access is required, and which adversarial transformations the method is designed to tolerate.
\item \textbf{A big data watermarking readiness framework.}
We define four deployment workloads: online generation, streaming detection, transformation pipelines, and ecosystem governance, and connect them to system-level requirements for scalability, reliability, robustness, utility, and governance.
\item \textbf{An evaluation blueprint and research agenda.} We identify workloads, metrics, and reporting practices needed to rigorously evaluate watermarking within realistic large-scale data pipelines rather than only on isolated model outputs.
\end{itemize}

\section{Watermarking: A Big Data Provenance Problem}
\subsection{From generated text to a data lifecycle}
A watermark is useful only if it remains detectable as generated content moves through its operational lifecycle. We consider five stages: \emph{generation}, an LLM produces watermarked text; \emph{transformation}, where the text may be paraphrased, translated, summarized, edited, mixed, or reformatted; \emph{storage and indexing}, where it enters repositories, search indexes, RAG corpora, or training datasets; \emph{verification}, where a detector estimates whether the watermark is present and, when supported, attributes its source; and \emph{governance}, where detection results inform auditing, access control, provenance tracking, or policy decisions. These stages may involve different systems and organizations, and content may pass through them repeatedly. Consequently, a watermark that is highly detectable immediately after generation may become increasingly weak or ambiguous after several downstream transformations.

This lifecycle perspective distinguishes watermarking from generic AI-text detection. While statistical detectors infer whether text resembles model output, watermarking deliberately embeds a signal during or after generation. This can strengthen provenance and attribution but depends on assumptions about control, keys, detector access, and downstream transformations. Watermarking is therefore best viewed as one component of a broader provenance architecture alongside metadata, cryptographic signatures, access logs, and governance policies.


Building on the five-V characterization of big data and prior research on watermarking reliability, robustness, and evaluation~\cite{liu2024survey,kirchenbauer2023watermark,pan2024markllm}, Table~\ref{tab:5v} synthesizes the corresponding requirements for watermark deployment. This mapping highlights why detection accuracy alone is insufficient for evaluating watermark effectiveness in large-scale data ecosystems.

\begin{table*}[t]
\caption{Mapping the 5Vs of big data to LLM-watermarking requirements.}
\label{tab:5v}
\centering
\small
\begin{tabularx}{\textwidth}{p{1.1cm} p{3.55cm} Y Y}
\toprule
\textbf{V} & \textbf{Big data condition} & \textbf{Watermarking requirement} & \textbf{Typical failure if ignored} \\
\midrule
Volume & Massive numbers of generated text & Low   false-positive rates; low latency and high throughput; scalable key management & A small false-positive rate can produce many false alarms at scale \\ \midrule
Velocity & Continuous model serving and streaming  & Low embedding latency; online or incremental detection; low memory usage & Expensive methods reduce serving throughput or cannot keep pace with incoming streams \\
\midrule
Variety & Languages, domains, code,  multi-stage pipelines & Cross-domain and cross-lingual robustness; explicit support for mixed   content & A detector performance degrades under distribution shifts or transformations   \\
\midrule
Veracity & Decisions depend on trustworthy provenance evidence & Calibrated scores,  robustness, attack resistance, uncertainty disclosure & Binary detection is over-interpreted despite weak evidence or adversarial manipulation \\
\midrule
Value & Provenance must justify operational and quality costs & Minimal  loss, manageable cost, interoperable verification, and governance alignment  & A  watermark may remain impractical due to utility loss or operational burden \\
\bottomrule
\end{tabularx}
\end{table*}

\subsection{Why scale changes watermark reliability}
Big data deployment changes how watermark reliability should be interpreted. A false-positive rate that appears negligible on a small benchmark can lead to many incorrect detections when a system processes millions of documents. Repeated content, correlated samples, and distribution shifts across models, domains, or languages can further amplify this problem \cite{cai2025optimal}. Watermark evaluation should therefore more accurately reflect population-scale operating conditions rather than rely only on accuracy measured on a narrow dataset

This issue is especially important when watermark detections support auditing, attribution, or policy decisions. Practical evaluations should report false-positive behavior at realistic operating points, detector calibration, uncertainty, and robustness under domain shifts and common text transformations. These measures provide a more meaningful view of veracity for large-scale provenance systems than  accuracy alone.

Scale also complicates attribution. Multi-bit and multi-key watermarking can associate generated content with a provider, user, session, or message~\cite{yoo2024advancing,wang2024building}. While this can strengthen accountability, it also increases the complexity of payload design, decoding, key management, and identity management. In large data ecosystems, watermarking must therefore be considered together with the infrastructure used to manage keys, detectors, identities, and provenance records.

\subsection{Scope and Systematization}

This paper systematizes representative LLM watermarking research covering generation-time token biasing, distortion-free or distribution-preserving sampling, semantic watermarking, post-processing methods, public verification, multi-bit attribution, attacks, and evaluation toolkits~\cite{kirchenbauer2023watermark,kuditipudi2023robust,christ2024undetectable,liu2023unforgeable,liu2024semantic,hou2024semstamp,chang2024postmark,pan2024markllm}. Our focus is on text generated by autoregressive LLMs and the operational properties that determine watermark reliability and scalability in large-scale data pipelines.

This work adopts a deployment-oriented systematization rather than an exhaustive systematic review. We prioritize representative peer-reviewed studies published between 2023 and 2026, alongside selected foundational works. Relevant studies were identified through searches of Google Scholar, IEEE Xplore, ACM Digital Library, and ACL Anthology using keywords such as ``LLM watermarking,'' ``text watermarking,'' ``watermark robustness,'' and ``LLM provenance.'' Studies are selected based on their relevance to watermark insertion, verification, robustness, security, and deployment efficiency. We exclude unrelated information-hiding techniques and methods outside LLM-generated text, except where they provide relevant methodological insights. Recent preprints are included selectively when addressing emerging deployment challenges not yet adequately covered by peer-reviewed literature.

Unlike existing surveys and deployment-oriented studies, which primarily examine watermarking mechanisms, security properties, and evaluation trade-offs~\cite{liu2024survey,dang2025sok,lalai2025intentions,piet2025markmywords}, our work emphasizes their implications for big data provenance. Specifically, we classify watermarking approaches along four deployment dimensions: insertion point, verification authority, operational state, and transformation threat model. We further connect these dimensions to the five Vs of big data and propose workload-specific readiness requirements for online generation, streaming detection, transformation pipelines, and ecosystem governance. This perspective links watermarking design choices to the scalability, reliability, interoperability, and governance requirements of large-scale provenance systems.

\section{A Deployment Taxonomy for LLM Watermarks}
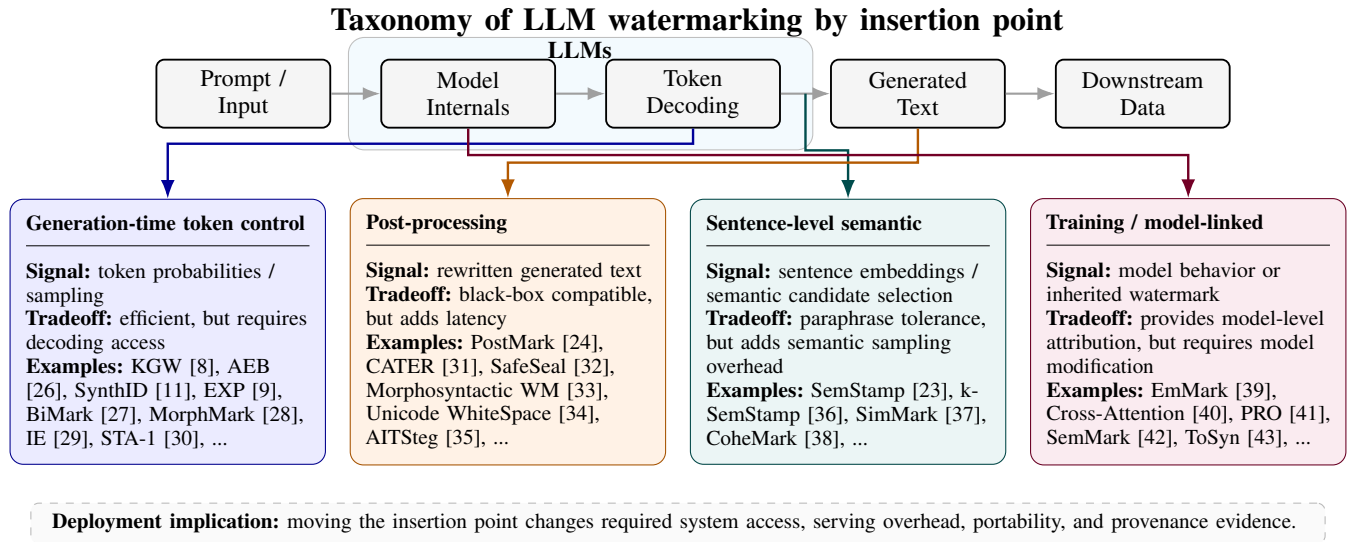
\begin{figure*}[t]
\centering
\begin{tikzpicture}[
    font=\small,
    >=Latex,
    stage/.style={
        draw,
        rounded corners=3pt,
        minimum height=0.9cm,
        minimum width=2.3cm,
        align=center,
        fill=gray!8,
        line width=0.7pt
    },
    method/.style={
        draw,
        rounded corners=5pt,
        text width=3.75cm, 
        minimum height=2.05cm,
        align=left,
        inner sep=6pt,
        font =\footnotesize 
    },
    token/.style={method, fill=blue!8, draw=blue!55!black},
    semantic/.style={method, fill=teal!8, draw=teal!55!black},
    post/.style={method, fill=orange!10, draw=orange!65!black},
    model/.style={method, fill=purple!8, draw=purple!55!black},
    arrow/.style={->, line width=0.8pt, draw=gray!75},
    link/.style={->, line width=0.9pt}
]

\node[font=\bfseries\large] (title) at (-2.0,3.75)
{LLM Watermarking Taxonomy: Insertion Points and Deployment Requirements};

\node[stage] (prompt) at (-8.0,2.8) {Prompt /\\Input};

\node[stage, right=0.65cm of prompt] (modelstage)
{Model\\Internals};

\node[stage, right=0.65cm of modelstage] (decode)
{Token\\Decoding};

\node[stage, right=0.65cm of decode] (output)
{Generated\\Text};

\node[stage, right=0.65cm of output] (downstream)
{Downstream\\Data};

\begin{pgfonlayer}{background}
\node[
    draw=gray!50,
    rounded corners=6pt,
    fill=cyan!4,
    inner xsep=12pt,
    inner ysep=8pt,
    fit=(modelstage)(decode)
] (llmbox) {};
\end{pgfonlayer}

\node[
    font=\bfseries\small,
    above=2.5pt of llmbox,
    xshift=-0.1mm,
    yshift=-5mm
] (llmlabel) {LLMs};

\draw[arrow] (prompt) -- (modelstage);
\draw[arrow] (modelstage) -- (decode);
\draw[arrow] (decode) -- (output);
\draw[arrow] (output) -- (downstream);
\node[token] (tokenwm) at (-9.0,-0.35) {
    \textbf{Generation-time token control}\\[-1mm]
    \rule{\linewidth}{0.3pt}\\[1mm]
    \textbf{Signal:} token probabilities / sampling\\
    \textbf{Tradeoff:} efficient, but requires decoding access \\
    \textbf{Examples:} KGW \cite{kirchenbauer2023watermark}, AEB \cite{cohen2025watermarking}, SynthID \cite{dathathri2024scalable}, EXP \cite{kuditipudi2023robust}, BiMark \cite{FengXiaoyan2025BUMW}, MorphMark \cite{wang2025morphmark}, IE \cite{gu2025invisible}, STA-1 \cite{mao2024watermark}, ... \\
};

\node[post] (postwm) at (-4.5,-0.35) {
    \textbf{Post-processing}\\[-1mm]
    \rule{\linewidth}{0.3pt}\\[1mm]
    \textbf{Signal:} rewritten generated text\\
    \textbf{Tradeoff:} black-box compatible, but adds latency \\
    \textbf{Examples:} PostMark \cite{chang2024postmark}, CATER \cite{he2022cater}, SafeSeal \cite{dang2026robustllmwatermarkingminimal}, Morphosyntactic WM \cite{meral2009natural}, Unicode WhiteSpace \cite{hellmeier2025hidden}, AITSteg \cite{ahvanooey2018aitsteg}, ...\\
};

\node[semantic] (semwm) at (0.0,-0.35) {
    \textbf{Sentence-level semantic}\\[-1mm]
    \rule{\linewidth}{0.3pt}\\[1mm]
    \textbf{Signal:} sentence embeddings / semantic candidate selection\\
    \textbf{Tradeoff:} paraphrase tolerance, but adds semantic sampling overhead \\
    \textbf{Examples:} SemStamp \cite{hou2024semstamp}, k-SemStamp \cite{hou2024ksemstamp}, SimMark \cite{dabiriaghdam2025simmark}, CoheMark \cite{ZhangJunyan2025CANS}, ... \\
};

\node[model] (modelwm) at (4.5,-0.35) {
    \textbf{Training / model-linked}\\[-1mm]
    \rule{\linewidth}{0.3pt}\\[1mm]
    \textbf{Signal:} model behavior or inherited watermark\\
    \textbf{Tradeoff:} provides model-level attribution, but requires
model modification \\
    \textbf{Examples:} EmMark \cite{zhang2024emmark}, Cross-Attention \cite{baldassini2024cross}, PRO \cite{xue2025pro}, SemMark \cite{LiHao2025FEtD}, ToSyn \cite{li2023protecting}, ...\\
};

\draw[link, blue!60!black]
    (decode.south) -- ++(0,-0.20) -| (tokenwm.north);

\draw[link, teal!60!black]
    ($(decode.east)!0.5!(output.west)$) -- ++(0,-0.75) -| (semwm.north);

\draw[link, orange!70!black]
    (output.south) -- ++(0,-0.45) -| (postwm.north);

\draw[link, purple!60!black]
    (modelstage.south) -- ++(0,-0.35) -| (modelwm.north);
\node[
    draw=gray!55,
    dashed,
    rounded corners=3pt,
    fill=gray!4,
    align=center,
    inner sep=2pt,
    inner xsep=8pt,
    inner ysep=4pt,
    font=\footnotesize,
    below=0.50cm of semwm,
    xshift=-2.25cm 
] (deployment) {
    \textbf{Deployment implication:}
    moving the insertion point changes required system access, serving overhead, portability, and provenance evidence.
};

\end{tikzpicture}

\caption{Deployment-oriented taxonomy of LLM watermarking by insertion point.
Generation-time methods intervene during decoding or semantic candidate
selection, post-processing methods modify completed outputs, and
training- or model-linked methods associate provenance with model behavior
or watermark inheritance across collections of outputs. Together, these classes expose distinct deployment tradeoffs across watermark lifecycles.}
\label{fig:watermark-insertion-taxonomy}
\end{figure*}

Algorithm names alone poorly predict deployment suitability, so we separate four complementary axes (Figure~\ref{fig:watermark-insertion-taxonomy}). 

\subsection{Insertion point: where the signal enters the pipeline?}
\textbf{Generation-time token control.} These methods embed the watermark while selecting each next token. Logit-based approaches bias or reweight the next-token distribution, as in KGW, Unigram, SIR, SemaMark, and adaptive watermarking \cite{kirchenbauer2023watermark,zhao2024provable,liu2024semantic,ren2024semamark,liu2024adaptive}. Other approaches modify the sampling rule, including exponential-minimum sampling, unbiased or distribution-preserving sampling \cite{kuditipudi2023robust,hu2024unbiased,wu2024dipmark,dathathri2024scalable}. These methods can be efficient because the signal is produced during decoding, but they require control of the generation stack and often weaken substantially on short or low-entropy text.


\textbf{Post-processing methods.} 
When logit access is unavailable, watermarking can modify generated text through lexical, syntactic, format, or semantic transformations. Lexical methods are the most common, using synonym substitution, phrase replacement, or controlled paraphrasing \cite{yang2023watermarking,qiang2023natural,xiang2024reversible,yang2022tracing,he2022lexical,chang2024postmark,dang2026robustllmwatermarkingminimal}. Syntactic methods alter grammatical structure or morphology \cite{topkara2006words,meral2009natural,xu2024robust}, while format-based methods use whitespace, Unicode, or other low-visibility markers \cite{hellmeier2025innamark, hellmeier2025hidden, sato2023embarrassingly}. Such methods are operationally attractive when the base model is closed, but they introduce additional transformations and may require another model or embedding service. Deployment must account for both extra latency and semantic drift.

\textbf{Sentence-level semantic methods.} Rather than encoding evidence in individual token choices, these methods operate on sentence-level semantic representations and select candidate sentences according to constraints in semantic space. SemStamp partitions sentence embeddings using locality-sensitive hashing, while k-SemStamp replaces this with k-means clustering \cite{hou2024semstamp,hou2024ksemstamp}. Related approaches use inter-sentence semantic similarity or cohesion-aware clustering and selection, as in SimMark and CoheMark \cite{dabiriaghdam2025simmark,ZhangJunyan2025CANS}.
Their advantage is improved tolerance to lexical editing; their cost is additional embedding, clustering, rejection sampling, or auxiliary-model computation. In a high-velocity system, this compute path must be evaluated as serving overhead rather than only as generation quality.

\textbf{Training- or model-linked provenance.} Some schemes embed provenance into model parameters, internal representations, or learned generation behavior rather than relying only on decoding-time token control. Related work includes EmMark, which embeds signatures into selected model weights \cite{zhang2024emmark}, Cross-Attention watermarking, which integrates a learned watermarking mechanism into the model \cite{baldassini2024cross}, PRO, which jointly trains watermark behavior with an open-source LLM \cite{xue2025pro}, and studies of watermark inheritance under knowledge distillation and model-level ownership verification \cite{pan2025canllm,shao2025explanation}. These methods can provide model-level provenance that accumulates across many outputs, which is particularly valuable for detecting extraction or imitation at big data scale. However, they may require model modification or internal access, and their provenance signal can be weakened by downstream operations such as fine-tuning, model merging, or adversarial distillation. Verification may also require aggregating evidence across many outputs, increasing data-retention and detection overhead compared with per-document watermarking.

\subsection{Verification authority: who is allowed to detect?}
\textbf{Private verification} uses a secret key or provider-controlled detector \cite{zhou2024bileve,LiZhuang2025BETW,li2024identifying,zhang2024remark}. It can reduce spoofing exposure but creates a trust bottleneck: external auditors must rely on the provider's interface or testimony. \textbf{Public verification} aims to let third parties validate provenance without revealing the generation secret \cite{liu2023unforgeable,Fairoze2023Publicly}. More recent work extends this direction using zero-knowledge proofs to demonstrate the correct execution of watermark detection while keeping the detection key secret \cite{duan2025pvmark}. The choice is therefore not merely cryptographic; it determines whether provenance evidence can be independently and transparently audited across organizational boundaries.

\subsection{Operational state: what must be available?}
Watermark schemes vary in the operational state required for generation and detection, including tokenizers, prompt or context, secret keys, reference sequences, model parameters, auxiliary encoders, and detector models \cite{kirchenbauer2023watermark,hou2024semstamp}. This state determines portability. Token-level detectors can be vulnerable when text modifications alter tokens and consequently disrupt keyed token relationships \cite{zhang2026character,zhang2023watermarks}, while semantic schemes may depend on a particular embedding encoder and semantic-space partition \cite{hou2024semstamp,ren2024semamark}. At big data scale, these dependencies must be versioned, stored, and discoverable so that long-lived content remains verifiable. Key rotation further introduces key-management and historical-key detection requirements \cite{aremu2026mitigating}. Model upgrades and key rotation therefore become data-governance problems, not only security-engineering details.


\subsection{Adversary model: what happens before detection?}
A meaningful robustness claim must state what happens between watermark embedding and detection. Common transformations include lexical edits, insertions, deletions, paraphrasing, summarization, translation, style transfer, human post-editing, document concatenation, and mixing with non-watermarked or human-written text. Adversarial variants include watermark removal, watermark stealing, and spoofing \cite{cheng2025revealing,pan2024markllm,krishna2024paraphrasing,dang2025delta,carlini2024stealing,jovanovic2024watermark,pasquini2025llmmap,an2026ditto,wu2024bypassing}. A system evaluated only on pristine generations therefore measures an upper bound on practical detectability.

\section{Scaling Watermarks Across the 5Vs}
\subsection{Volume: false positives, aggregation, and population risk}
Watermark papers often emphasize true-positive rate at a chosen false-positive rate. For large-scale deployment, the false-positive operating point is itself a systems parameter. A platform scanning hundreds of millions of items cannot interpret a one-percent false-positive rate in the same way as a laboratory benchmark. Even substantially smaller rates can create large absolute review queues at population scale. Importantly, detection precision also depends on the prevalence of watermarked content. For illustration, at a watermark prevalence of 0.1\%, a true-positive rate of 90\%, and a false-positive rate of 0.1\%, screening one million documents would produce approximately 900 true positives and 999 false positives, yielding a precision of only 47.4\%. Thus, a low false-positive rate does not necessarily ensure reliable attribution in populations where watermarked content is rare. Moreover, false positives are not uniformly costly in real deployments: an internal telemetry flag, a content label, an enforcement action, and a legal attribution require different evidentiary thresholds.

Volume can also help. When the task is model-extraction detection, evidence may be aggregated across many outputs. A weak per-sample signal can become statistically meaningful at dataset scale, provided samples are sufficiently independent and the detector accounts for repeated prompts, duplicated content, and correlated transformations. This suggests two distinct benchmark modes: \emph{item-level provenance} and \emph{collection-level provenance}. They should not be reported interchangeably.

Multi-use watermarking introduces another scale effect. If many keys are accepted, the probability that natural text aligns with at least one detectable pattern can increase; recent work explicitly studies this false-detection problem \cite{fu2025multi}. Deployment should therefore report not only single-key FPR but also FPR under the intended number of active keys, users, or tenants. This is   important in multi-tenant systems where verification must scale without inflating aggregate false-alarm rates.

\subsection{Velocity: serving latency and streaming detection}
For online generation, watermark embedding sits on the critical serving path. Token-bias methods can add modest per-token operations, whereas rejection sampling, multiple candidate generations, or semantic post-processing can add larger latency and compute cost \cite{hou2024semstamp,giboulot2024watermax,chang2024postmark}. A method that preserves quality but doubles generation cost may be unacceptable for a high-throughput service in large-scale production. Papers should consequently report throughput (tokens/s or documents/s), p50/p95 latency, additional model calls, GPU/CPU utilization where relevant, and detector throughput.

Detection can also be streaming. Many token-level detectors accumulate a statistic over a sequence, enabling an online score that strengthens as text arrives. This creates an opportunity for early stopping: once evidence is decisive, the detector need not process the entire document. Conversely, semantic methods may require a complete sentence or document before detection can proceed. These implementation differences matter in real-time moderation, indexing, and ingestion systems. 

\subsection{Variety: cross-domain, multilingual,  and structured data}
Text ecosystems are heterogeneous. Watermarks are exposed to different tokenizers, model families, languages, writing styles, code, tables, and mixed human/AI authorship. Translation is particularly challenging because it can change both lexical choices and sentence structure \cite{he2024can}. Semantic methods may improve invariance but inherit the cross-lingual quality and domain biases of their representation models. Code introduces another distribution with stricter syntactic constraints and different notions of acceptable perturbation.

Variety also arises from pipeline composition. A watermarked answer may be summarized by another LLM, inserted into a report, translated, and later used as RAG context. Each stage can reduce signal-to-noise. Robustness should therefore be evaluated not only against one attack at a time but against realistic chains of transformations. Mixed-authorship evaluation is equally important: detectors should report how scores behave when only a fraction of a document is watermarked.

\subsection{Veracity: calibrated evidence under attacks and shift}
Watermark veracity has two components. First, the statistical detector must be calibrated: a score should correspond to a controlled error rate on relevant non-watermarked data. Second, the watermark should resist adversarial removal and spoofing. Paraphrasing and lexical edits can substantially reduce detection for token-level schemes \cite{pang2024no}; watermark stealing can reveal exploitable structure from black-box queries \cite{jovanovic2024watermark}; and impossibility results caution against assuming a universally strong watermark under unconstrained transformations \cite{zhang2023watermarks}.

These results argue against binary claims such as ``watermarked'' or ``non-watermarked'' without context. A deployable detector should expose a score, threshold, calibration domain, expected FPR, supported text length, version identifier when appropriate, and known failure modes. In high-stakes settings, the detector should be treated as one piece of provenance evidence rather than definitive proof of authorship or intent.

\subsection{Value: quality, cost, and adoption incentives}
The value of watermarking depends on whether provenance benefits outweigh costs. Biasing token probabilities can alter fluency or downstream task performance, and stronger signals can trade off with quality \cite{giboulot2024watermax,molenda2024waterjudge,piet2025markmywords}. Semantic approaches may preserve meaning under attacks but consume additional inference resources. Public verification can improve ecosystem trust but increase engineering and cryptographic complexity. A complete evaluation should therefore include quality, latency, throughput, energy/compute proxy, and integration cost alongside detection accuracy under realistic deployment constraints.

Adoption is also strategic. If users can migrate to a non-watermarked competitor, a provider may bear quality and cost penalties without receiving the full social benefit of provenance. Recent position work emphasizes stakeholder incentives as a barrier to practical adoption \cite{liu2026incentives}. This is especially relevant in big data ecosystems where content crosses organizational boundaries and the party paying the embedding cost may not be the party receiving the verification benefit.

\section{Security, Trust, and Governance in Big Data  }

Deploying LLM watermarking at scale introduces challenges beyond detectability and computational efficiency. In big data ecosystems, provenance signals may be exposed to adversarial removal, stealing, or spoofing \cite{pang2024no,jovanovic2024watermark}, processed across heterogeneous users and domains~\cite{cohen2025watermarking,fu2025multi}, and verified by organizations with different levels of access and trust \cite{duan2025pvmark,luan2026vow}. As a result, watermarking must provide not only robust detection, but also secure attribution, fair and auditable decisions, privacy-aware identity management, and reproducible governance over keys and detectors \cite{luan2026vow,aremu2026watermarking,aremu2026mitigating}. These requirements become increasingly important at operational scale as watermarked content moves across organizational boundaries and persists over long data lifecycles, where watermark signals may be aggregated and linked across outputs over time~\cite{aremu2026watermarking}. This section examines these concerns through three complementary perspectives: removal, spoofing, and stealing attacks; privacy, fairness, and accountability; and governance of watermark keys, detectors, and versions.


\subsection{Removal, spoofing, and stealing attacks}

\textbf{Removal attacks} attempt to weaken or erase a watermark while preserving the semantic utility of the text. Common transformations include synonym substitution, paraphrasing, back-translation, summarization, sentence restructuring, and LLM-based rewriting \cite{cheng2025revealing,krishna2024paraphrasing,zhang2023watermarks}. Their effectiveness depends strongly on the watermark construction and detector design \cite{dang2025sok,liu2024survey}. From a big data perspective, removal is especially important because generated content may undergo multiple transformations as it passes through retrieval systems, summarization pipelines, data-cleaning stages, or downstream platforms. Robustness should therefore reflect how provenance signals degrade across repeated or chained transformations rather than only after a single isolated attack.

\textbf{Spoofing attacks} reverse this objective by causing non-watermarked or attacker-authored text to be falsely attributed to a watermarked model \cite{an2026ditto,gu2023learnability,jovanovic2024watermark,wu2024bypassing}. Instead of removing provenance evidence, spoofing introduces false provenance into the data ecosystem. At scale, such errors can rapidly and widely propagate through repositories, moderation systems, audit logs, and automated policy-enforcement pipelines. This makes spoofing particularly consequential when watermark detections are used to support model attribution, ownership claims, compliance decisions, or downstream filtering.

\textbf{Watermark stealing} attempts to infer the watermark structure or detection behavior from observed outputs, which can later support watermark removal or spoofing attacks \cite{jovanovic2024watermark,gu2023learnability}. 

\noindent \textbf{Model stealing}, in contrast, repeatedly queries a protected model and uses the collected outputs to construct or approximate a surrogate model \cite{dang2025delta,carlini2024stealing}. In this setting, watermarking can provide evidence of unauthorized model use by aggregating signals across many suspicious outputs. This aggregate view is particularly relevant to big data systems, where extraction attacks may involve large numbers of queries and provenance evidence can be evaluated across the resulting dataset.




\subsection{Privacy, fairness, and accountability}
\textbf{Privacy} becomes important when watermarks carry more than a binary provenance signal. Multi-bit schemes can encode identifiers associated with a user, session, model instance, or organization \cite{yoo2024advancing,wang2024building}. While this enables   attribution, it can also create persistent links among generated records and introduce privacy, retention, and re-identification risks. In a big data deployment, providers must therefore define what identity information is encoded or mapped to a watermark, who is authorized to resolve it, how long keys and identity mappings are retained, and how compromised or expired identifiers are revoked.  When millions of 
outputs are stored across multiple systems, these choices become part of the broader data-governance and access-control architecture rather than only properties of the watermark algorithm.

\textbf{Fairness} concerns whether watermark errors and quality effects are distributed consistently across heterogeneous data populations. Performance can vary across languages, domains, entropy levels, and text lengths~\cite{he2024can,lee2023wrote,tu2024waterbench}. In big data systems, even small  disparities can scale into large numbers of unevenly distributed false attributions or missed detections. Deployment should report stratified false-positive and false-negative rates, calibration, and quality effects across relevant dimensions rather than relying only on aggregate performance.

\textbf{Accountability} requires watermark evidence to remain consistently auditable as it moves through downstream systems. Because detection depends on calibration, key state, detector version, and the assumed threat model, a positive watermark result should be treated as evidence with an associated uncertainty rather than proof. Providers should record the detector version, calibration data, supported transformations, key-management assumptions, and known failure modes needed to reproduce a decision. Publicly verifiable watermarking and zero-knowledge-based verification can further reduce reliance on provider-controlled detection by allowing independent parties to validate provenance claims without exposing sensitive generation or detection secrets \cite{liu2023unforgeable,duan2025pvmark,Fairoze2023Publicly}. At big data scale, such metadata and verification records become part of the provenance infrastructure needed to support secure, scalable, and reproducible audits across organizations and over time.

\subsection{Governance of keys, detectors, and versions}


Key and detector governance become a lifecycle problem once watermarking is deployed across many users, models, and datasets. Private-key schemes require key generation, storage, rotation, access control, revocation, and mappings between identities and historical keys; multi-user watermarking makes these dependencies particularly important for reliable attribution \cite{cohen2025watermarking}. Detector compatibility must also be maintained as models and tokenizers change, since changes in tokenization can disrupt token-level watermark signals \cite{zhang2026character}. Semantic schemes introduce additional versioned dependencies such as embedding models and semantic-space partitions \cite{hou2024semstamp}. Publicly detectable or verifiable schemes reduce reliance on provider-controlled detection, but change the security model and must preserve resistance to forgery or removal \cite{Fairoze2023Publicly,duan2025pvmark}.

From a big data perspective, watermark configuration should therefore be treated as persistent, auditable, versioned managed provenance metadata. Each deployed version should record a configuration identifier, detector and key version, validity period, supported model and tokenizer versions, calibration information, and key-management policy. Maintaining these records allows historical content to be interpreted with the configuration under which it was generated rather than with only the current detector. Watermark versioning thus becomes part of the data-governance infrastructure required for reproducible provenance across large collections and long data lifecycles.

\section{Big Data Watermarking Readiness: Evaluation and Deployment Guidance}

Existing toolkits and benchmarks have made watermark evaluation more systematic \cite{pan2024markllm,piet2025markmywords}. For big data readiness, however, evaluation should reflect both system-scale workloads and realistic deployment settings. In particular, a method that performs well on a fixed benchmark may still be difficult to operate under continuous generation, heterogeneous data, repeated transformations, or multi-tenant verification. To address this gap, we define a \textbf{Big Data Watermarking Readiness framework} organized around four deployment workloads: \textbf{W1) online generation, W2) streaming detection, W3) transformation pipelines, and W4) ecosystem and governance}. Each workload captures a different stage or requirement of large-scale watermark deployment and identifies the metrics needed to evaluate whether a watermark is practical beyond isolated benchmark settings. Table~\ref{tab:benchmark} summarizes the minimum reporting requirements for these workloads.

\begin{table*}[t]
\caption{Minimum reporting requirements across the four Big Data Watermarking Readiness workloads.}
\label{tab:benchmark}
\centering
\small
\begin{tabularx}{\textwidth}{p{3.5cm} Y Y}
\toprule
\textbf{Workload} & \textbf{Report} & \textbf{Purpose} \\
\midrule

Online generation &
Throughput, p50/p95 latency, extra model calls, watermark strength, output quality &
Measures serving overhead and quality cost \\

Streaming detection &
Detection throughput, memory, detection latency, TPR/FPR, diverse non-watermarked data &
Measures scalable and reliable detection \\

Transformation pipelines &
Detection and text utility after individual and sequential transformations &
Measures whether provenance survives downstream processing \\

Ecosystem and governance &
Multi-key FPR, version compatibility, key rotation/revocation,
verification authority &
Measures reproducibility and long-term auditability \\

\bottomrule
\end{tabularx}
\end{table*}

\textbf{W1: Online generation.} Production LLM services require watermarking mechanisms with minimal serving overhead, since watermark computation lies on the latency-critical generation path \cite{dathathri2024scalable}. Evaluation should measure embedding cost under realistic prompt and output-length distributions, reporting throughput, latency distributions, additional model calls, and output quality relative to the non-watermarked baseline \cite{tu2024waterbench,kwon2023efficient}. Watermark strength should also be explicitly reported, since stronger perturbations may improve detectability or robustness at the cost of quality and serving efficiency. This workload represents \emph{provider-controlled API provenance}, where watermarking is integrated directly into the generation pipeline. Evaluation should further consider workload variability across short and long outputs and different concurrency levels, which can substantially affect serving efficiency \cite{kwon2023efficient,agrawal2023sarathi}. Finally, average throughput alone is insufficient: watermarking should not introduce unstable tail latency under peak load, since production serving systems must satisfy latency constraints across requests \cite{zhong2024distserve}. Methods that significantly reduce throughput, increase tail latency, or degrade output quality may therefore be unsuitable for big data deployment.

\textbf{W2: Streaming detection.}
Large platforms may need to verify high-volume, heterogeneous text streams in near real time, making detector efficiency and false-positive control system-level requirements \cite{dathathri2024scalable,akidau2015dataflow}. Evaluation should therefore process mixtures of watermarked and non-watermarked documents at production-like rates and report detection throughput (documents/s), memory use, detection latency, and ROC or precision-recall performance at deployment-relevant thresholds \cite{kirchenbauer2023reliability,tu2024waterbench}. Non-watermarked data should include human-written text, ordinary outputs from different LLMs, and content spanning multiple domains, document lengths, programming code, and languages, since watermark performance can vary substantially across these settings \cite{he2024can,lee2023wrote,tu2024waterbench}. Long documents may contain only a small portion of watermarked text, especially when content is copied, combined, or rewritten. Detection should therefore identify whether watermarked content is present without requiring expensive processing of the entire document \cite{pan2025waterseeker,zhao2025efficiently}. Evaluation should also consider different document lengths and changes in request volume, since both can affect detection speed and system throughput.


\textbf{W3: Transformation pipelines.} Generated text may be modified before verification through paraphrasing, translation, summarization, truncation, formatting changes, document mixing, or LLM-based rewriting. Prior work shows that paraphrasing, translation, and mixing with other text can substantially weaken watermark signals \cite{kirchenbauer2023reliability,he2024can,pan2024markllm,zhang2023watermarks}. 
Evaluation should therefore measure both \emph{watermark detectability} and \emph{text utility} after each transformation, since stronger modifications may remove the watermark but also reduce semantic similarity, readability, or downstream-task performance. This trade-off is important when comparing robustness across methods: a watermark should not be considered weak simply because an attack succeeds only by substantially damaging the text. Comparisons should also control for watermark strength \cite{tu2024waterbench} and test both individual and sequential transformations. From a big data perspective, this evaluation shows whether provenance and useful content can survive as text moves through multiple processing stages over its lifecycle.

\textbf{W4: Ecosystem and governance.}
Real deployments may involve many users, watermark keys, model versions, detector versions, and verification organizations. Evaluation should test how false-positive rates change as the number of active users or keys grows \cite{cohen2025watermarking,fu2025multi}. It should also examine version compatibility, including whether previously generated content remains verifiable after keys, models, tokenizers, or detectors are updated. Public and private verification workflows should be evaluated when provenance must be shared across organizations with different levels of trust \cite{liu2023unforgeable,Fairoze2023Publicly,duan2025pvmark}. Multi-user watermarks can support user-level attribution, while model-protection schemes can aggregate evidence across multiple outputs to support model-ownership or imitation detection \cite{cohen2025watermarking,pang2025modelshield}. From a big data perspective, keys, configurations, detector versions, validity periods, and verification records should therefore be managed as provenance metadata so that decisions remain reproducible across large datasets and long data lifecycles across evolving organizational boundaries.

These four workloads show that watermark deployment readiness cannot be determined by detection accuracy alone. Online generation requires low serving overhead and preserved output quality; streaming detection requires scalable processing and reliable false-positive control; transformation pipelines require both watermark detectability and text utility to survive downstream processing; and ecosystem deployment requires reliable operation across users, keys, models, domains, and verification authorities. Table~\ref{tab:benchmark} summarizes the minimum reporting requirements for evaluating these conditions. Overall, the appropriate watermark design depends on the intended workload, generation access, verification model, expected transformations, acceptable error rates, and operational constraints. A deployable watermark should therefore be evaluated as provenance infrastructure that remains accurate, efficient, useful, and auditable throughout the big data lifecycle.

\section{Open Research Directions and Limitations}

Despite substantial progress, several challenges remain before LLM watermarking can operate as reliable provenance infrastructure in large-scale and evolving data ecosystems.

\textbf{Population-scale and continual calibration.}
Watermark detectors must maintain predictable false-positive behavior as data distributions, model versions, and watermark keys change over time. Calibration that is valid for one model or benchmark may not remain valid after deployment across new domains, languages, or users. Repeated or correlated content and multi-key verification can further complicate statistical assumptions \cite{cai2025optimal,fu2025multi}. A key research question is how to maintain reliable error guarantees under continuous model and data drift while preserving comparability with historical provenance decisions. Calibration should be treated as an ongoing operational process rather than a one-time benchmark result. Future systems should therefore support adaptive recalibration while preserving stable decision thresholds and auditability across deployment periods.

\textbf{Compositional robustness.}
Most robustness evaluations study transformations individually, while real data pipelines may apply several transformations in sequence. Watermarked content can be paraphrased, translated, summarized, reformatted, mixed with other text, and rewritten again before verification \cite{kirchenbauer2023reliability,he2024can}. The effects of these transformations may compound, causing watermark evidence to decay differently from what single-transformation results suggest. Future work should measure how detection and text utility change across transformation sequences, determine whether robustness to individual transformations predicts robustness to their composition, and investigate whether provenance confidence can be updated as content moves through successive processing stages.

\textbf{Scalable and distributed verification.}
High-volume platforms may need to verify watermark evidence across continuous streams, large repositories, or distributed processing systems. Existing work has begun to improve efficient detection of partially watermarked content in large documents~ \cite{pan2025waterseeker,zhao2025efficiently}, but less is known about how detection statistics behave when verification is parallelized or distributed across machines. Future systems should investigate incremental processing, early stopping, distributed aggregation, and resource-aware detection while preserving false-positive guarantees. The central challenge is to improve throughput and time-to-decision without changing the statistical meaning of the detector output.

\textbf{Heterogeneous and cross-domain provenance.}
Practical data ecosystems contain multilingual text, code, structured documents, different model families, and mixed human-AI content. Watermark performance can vary substantially across languages, domains, entropy levels, and text lengths \cite{he2024can,lee2023wrote,tu2024waterbench}. Future work should move beyond English prose and determine whether watermark signals remain robust, reliable, and comparable across heterogeneous content types, tokenizers, and generation environments. Code refactoring, translation, format conversion, and domain-specific rewriting introduce particularly challenging transformations. More broadly, extending these principles to multimodal content remains an important direction for provenance systems that increasingly combine text with images, audio, and other modalities.

\textbf{Interoperable governance.}
Provenance may need to be checked by providers, platforms, auditors, and other organizations that do not share the same keys or level of trust. Publicly verifiable and zero-knowledge-based methods provide useful support for independent verification \cite{liu2023unforgeable,Fairoze2023Publicly,duan2025pvmark}, but cross-organization use remains challenging. Future systems need common ways to represent watermark identifiers, detector scores, confidence levels, verification authority, key revocation, and detector information. They should also define how provenance evidence can be shared and interpreted consistently across organizations while protecting sensitive keys and resisting watermark removal, spoofing, and stealing.

\textbf{Longitudinal deployment benchmarks.}
Current benchmarks largely provide static evaluations of watermark algorithms under fixed models and experimental settings \cite{pan2024markllm,tu2024waterbench}. Big data provenance systems, however, evolve over time. Future benchmarks should evaluate whether previously generated content remains verifiable as models and tokenizers are upgraded, keys are rotated or revoked, detectors are updated or recalibrated, and data distributions shift. They should jointly measure detection, utility, latency, throughput, transformation robustness, and compatibility across versions. Such long-term evaluation would show whether watermark evidence remains reliable and reproducible throughout the lifecycle of large-scale data rather than only under the original generation setting.

\textbf{Limitations.}
This systematization has several limitations. First, our deployment-oriented taxonomy emphasizes system and operational properties rather than providing an exhaustive comparison of every watermarking algorithm or implementation detail. Second, results across studies remain difficult to compare because evaluations use different models, decoding settings, watermark strengths, text lengths, datasets, attacks, and operating thresholds. This limits direct quantitative comparison and motivates more standardized reporting and benchmarking. Third, the proposed 5V framework provides a systems-oriented lens for organizing deployment requirements, but its usefulness should be further validated through common benchmarks and real-world deployments spanning multiple models, languages, domains, and transformation pipelines. Finally, our primary scope is textual watermarking for LLM-generated content; multimodal provenance, metadata-based provenance, and cryptographic content-credential systems are considered only where they directly interact with text watermarking. Future work can extend this framework to broader provenance architectures that combine multiple complementary signals.

\section{Conclusion}
LLM watermarking should be treated as provenance infrastructure for large-scale data systems rather than only as a mechanism for detecting individual model outputs. Through the 5V perspective, this systematization shows that deployment readiness depends on more than detection accuracy: Volume requires population-scale false-positive control, Velocity requires efficient generation and detection, Variety requires robustness across heterogeneous data and transformations, Veracity requires calibrated and attack-resistant evidence, and Value requires that provenance benefits justify quality, computational, and governance costs. Our deployment-oriented taxonomy further shows that insertion point, verification authority, operational state, and adversary model directly shape where and how a watermark can be used in real-world deployments.

To translate these requirements into practice, we organize evaluation around online generation, streaming detection, transformation pipelines, and ecosystem governance. Across these workloads, a deployable watermark must remain detectable without unacceptable loss of text utility, scale to realistic serving and verification workloads, and support reproducible decisions as models, keys, detectors, and data evolve over time. In general, no single watermarking family satisfies all deployment requirements. Progress toward trustworthy LLM provenance will therefore require joint consideration of statistical reliability, system efficiency, robustness, interoperability, and long-term governance throughout the big data lifecycle.

\section*{Acknowledgment}
This work was supported by the State University of New York at Albany, under IFR 940008-20 and the Chancellor's Summer Research Excellence Fund. We thank Eoin Carr for his valuable support and helpful discussions for this project.

\bibliographystyle{IEEEtran}
\bibliography{survey}

@inproceedings{carlini2024stealing,
title = {Stealing part of a production language model},
author = {Carlini, Nicholas and Paleka, Daniel and Dvijotham, Krishnamurthy (Dj) and Steinke, Thomas and Hayase, Jonathan and Cooper, A. Feder and Lee, Katherine and  others},
year = {2024},
booktitle = {ICML},
articleno = {221},
numpages = {26},
location = {Vienna, Austria},
}

@misc{gemini, title={Google Gemini}, journal={Google}, author={Google},
 howpublished = {https://bard.google.com/chat/}, year={2026}}

@article{fairoze2023publicly,
         author={Fairoze, Jaiden and Garg, Sanjam and Jha, Somesh and Mahloujifar, Saeed and Mahmoody, Mohammad and Wang, Mingyuan},
         title={Publicly-Detectable Watermarking for Language Models},
         volume={1},
         number={4},
         year={2025},
         journal={{IACR} Communications in Cryptology},
 }

@article{kuditipudi2023robust,
  title={Robust distortion-free watermarks for language models},
  author={Kuditipudi, Rohith and Thickstun, John and Hashimoto, Tatsunori and Liang, Percy},
  journal={TMLR},
  year={2023}
}

@inproceedings{pan2024markllm,
  title={MarkLLM: An Open-Source Toolkit for LLM Watermarking},
    author = "Pan, Leyi  and
      Liu, Aiwei  and
      He, Zhiwei  and
      Gao, Zitian  and
      Zhao, Xuandong  and
      Lu, Yijian  and
      Zhou, Binglin  and
     others",
    booktitle = "EMNLP",
    year = "2024",
}

@inproceedings{an2026ditto,
  title={Ditto: A spoofing attack framework on watermarked llms via knowledge distillation},
  author={An, Hyeseon and Park, Shinwoo and Woo, Suyeon and Han, Yo-Sub},
  booktitle={Proceedings of the 19th Conference of the European Chapter of the Association for Computational Linguistics (Volume 1: Long Papers)},
  year={2026}
}

@article{gu2023learnability,
  title={On the learnability of watermarks for language models},
  author={Gu, Chenchen and Li, Xiang Lisa and Liang, Percy and Hashimoto, Tatsunori},
  journal={ICLR},
  year={2024}
}

@inproceedings{lee2023wrote,
  title={Who wrote this code? watermarking for code generation},
  author={Lee, Taehyun and Hong, Seokhee and Ahn, Jaewoo and Hong, Ilgee and Lee, Hwaran and Yun, Sangdoo and Shin, Jamin and Kim, Gunhee},
  booktitle={ACL},
  year={2024}
}

@article{mao2024watermark,
  title={A Watermark for Low-entropy and Unbiased Generation in Large Language Models},
  author={Mao, Minjia and Wei, Dongjun and Chen, Zeyu and Fang, Xiao and Chau, Michael},
  journal={arXiv},
  year={2024}
}

@inproceedings{giboulot2024watermax,
  title={WaterMax: breaking the LLM watermark detectability-robustness-quality trade-off},
  author={Giboulot, Eva and Teddy, Furon},
  booktitle={NeurIPS},
  year={2024}
}

@article{xiang2024reversible,
  title={A reversible natural language watermarking for sensitive information protection},
author = {Xiang, Lingyun and Liu, Yangfan and Yang, Zhongliang},
  journal={Information Processing \& Management},
  year={2024},
 }

@inproceedings{yang2022tracing,
  title={Tracing text provenance via context-aware lexical substitution},
  author={Yang, Xi and Zhang, Jie and Chen, Kejiang and Zhang, Weiming and Ma, Zehua and others},
  booktitle={AAAI},
  year={2022}
}

@inproceedings{topkara2006words,
  title={Words are not enough: sentence level natural language watermarking},
  author={Topkara, Mercan and Topkara, Umut and Atallah, Mikhail J},
  booktitle={Proceedings of the 4th ACM international workshop on Contents protection and security},
  pages={37--46},
  year={2006}
}

@article{qiang2023natural,
  title={Natural language watermarking via paraphraser-based lexical substitution},
  author={Qiang, Jipeng and Zhu, Shiyu and Li, Yun and Zhu, Yi and Yuan, Yunhao and Wu, Xindong},
  journal={Artificial Intelligence},
   pages={103859},
  year={2023},
 }

@InProceedings{wang2025morphmark,
  title={MorphMark: Flexible Adaptive Watermarking for Large Language Models},
  author={Wang, Zongqi and Gu, Tianle and Wu, Baoyuan and Yang, Yujiu},
  booktitle={ACL},
  year={2025}
}

@inproceedings{liu2024adaptive,
  title={Adaptive text watermark for large language models},
  author={Liu, Yepeng and Bu, Yuheng},
  booktitle={ICML},
  year={2024}
}

@misc{dang2026robustllmwatermarkingminimal,
  title={Robust LLM Watermarking with Minimal Semantic Distortion for IP Protection},
  author={Dang, Kieu and Lai, Phung and Phan, NhatHai and Shen, Yelong and Jin, Ruoming},
  journal={arXiv},
  year={2026}
}

@inproceedings{ren2024semamark,
  title={SemaMark: A Semantics-Based Watermark Robust to Paraphrasing},
      author = "Ren, Jie  and
      Xu, Han  and
      Liu, Yiding  and
      Cui, Yingqian  and
      Wang, Shuaiqiang  and
      Yin, Dawei  and
      Tang, Jiliang",
  booktitle={Findings of the Association for Computational Linguistics: NAACL 2024},
  year={2024}
}

@inproceedings{tu2024waterbench,
  title={Waterbench: Towards holistic evaluation of watermarks for large language models},
  author={Tu Shangqing and Yuliang Sun and Yushi Bai and Jifan Yu and Lei Hou and Juanzi Li},
  booktitle={ACL},
  year={2024}
}

@inproceedings{piet2025markmywords,
  title={MARKMyWORDS: Analyzing and Evaluating Language Model Watermarks},
  author={Piet, Julien and Sitawarin, Chawin and Fang, Vivian and Mu, Norman and Wagner, David},
  booktitle={SaTML},
  year={2025},
}

@inproceedings{liu2024semantic,
  title     = {A Semantic Invariant Robust Watermark for Large Language Models},
  author    = {Liu, Aohan and Pan, Liangming and Hu, Xiaojun and Meng, Shiqi and Wen, Lin},
  booktitle = {International Conference on Learning Representations (ICLR)},
  year      = {2024}
}

@inproceedings{hou2024ksemstamp,
  title     = {k-SemStamp: A Clustering-Based Semantic Watermark for Detection of Machine-Generated Text},
    author={Hou, Abe and Zhang, Jingyu and Wang, Yichen and Khashabi, Daniel and He, Tianxing},
  booktitle = {ACL},
  year      = {2024}
}

@inproceedings{he2022lexical,
  title     = {Protecting Intellectual Property of Language Generation APIs with Lexical Watermark},
  author    = {He, Xiaojun and Xu, Qiang and Lyu, Lingjuan and Wu, Fei and Wang, Cong},
  booktitle = {Proceedings of the AAAI Conference on Artificial Intelligence},
  year      = {2022}
}

@inproceedings{kirchenbauer2023watermark,
  title={A watermark for large language models},
  author={Kirchenbauer, John and Geiping, Jonas and Wen, Yuxin and Katz, Jonathan and Miers, Ian and Goldstein, Tom},
  booktitle={International Conference on Machine Learning},
  pages={17061--17084},
  year={2023},
  organization={PMLR}
}

@inproceedings{lalai2025intentions,
  title={From intentions to techniques: A comprehensive taxonomy and challenges in text watermarking for large language models},
  author={Lalai, Harsh Nishant and Ramakrishnan, Aashish Anantha and Shah, Raj Sanjay and Lee, Dongwon},
  booktitle={NAACL},
  year={2025}
}

@article{dang2025sok,
  title={SoK: Are Watermarks in LLMs Ready for Deployment?},
  author={Dang, Kieu and Lai, Phung and Phan, NhatHai and Shen, Yelong and Jin, Ruoming and Khreishah, Abdallah and Thai, My T},
  journal={arXiv},
  year={2025}
}

@article{zhang2023watermarks,
  title={Watermarks in the sand: Impossibility of strong watermarking for generative models},
  author={Zhang, Hanlin and Edelman, Benjamin L and Francati, Danilo and Venturi, Daniele and Ateniese, Giuseppe and Barak, Boaz},
  journal={ICML},
  year={2024}
}

@misc{OpenAI,
  title        = {OpenAI API},
  author       = {{OpenAI}},
  year         = {2026},
  url          = {https://openai.com/index/openai-api/},
  note         = {Accessed: 2026-02-03}
}

@inproceedings{zhou2024bileve,
author = {Zhou, Tong and Zhao, Xuandong and Xu, Xiaolin and Ren, Shaolei},
title = {Bileve: securing text provenance in large language models against spoofing with bi-level signature},
year = {2024},
booktitle = {Proceedings of the 38th International Conference on Neural Information Processing Systems},
articleno = {1783},
numpages = {22},
series = {NIPS '24}
}

@article{dathathri2024scalable,
  title={Scalable watermarking for identifying large language model outputs},
  author={Dathathri, Sumanth and See, Abigail and Ghaisas, Sumedh and Huang, Po-Sen and McAdam, Rob and Welbl, Johannes and Bachani, Vandana and Kaskasoli, Alex and Stanforth, Robert and others},
  journal={Nature},
  year={2024},
}

@inproceedings{wu2024dipmark,
  title={Dipmark: A stealthy, efficient and resilient watermark for large language models},
  author={Wu, Yihan and Hu, Zhengmian and Zhang, Hongyang and Huang, Heng},
  booktitle={ICML},
  year={2024}
}

@article{krishna2024paraphrasing,
  title={Paraphrasing evades detectors of ai-generated text, but retrieval is an effective defense},
  author={Krishna, Kalpesh and Song, Yixiao and Karpinska, Marzena and Wieting, John and Iyyer, Mohit},
  journal={NeurIPS},
  volume={36},
  year={2024}
}

@inproceedings{zhang2024remark,
  title={{REMARK-LLM}: A Robust and Efficient Watermarking Framework for Generative LLMs},
  author={Zhang, Ruisi and Hussain, Shehzeen Samarah and Neekhara, Paarth and Koushanfar, Farinaz},
  booktitle={USENIX},
  year={2024}
}

@inproceedings{christ2024undetectable,
  title={Undetectable watermarks for language models},
  author={Christ, Miranda and Gunn, Sam and Zamir, Or},
  booktitle={The Thirty Seventh Annual Conference on Learning Theory},
  year={2024},
  organization={PMLR}
}

@inproceedings{he2024can,
  title={Can watermarks survive translation? on the cross-lingual consistency of text watermark for llms},
  author={He, Zhiwei and Zhou, Binglin and Hao, Hongkun and Liu, Aiwei and Wang, Xing and others},
  booktitle={ACL},
  year={2024}
}

@inproceedings{li2023protecting,
  title={Protecting intellectual property of large language model-based code generation apis via watermarks},
  author={Li, Zongjie and Wang, Chaozheng and Wang, Shuai and Gao, Cuiyun},
  booktitle={ACM SIGSAC},
  pages={2336--2350},
  year={2023}
}

@inproceedings{liu2023unforgeable,
  title={An unforgeable publicly verifiable watermark for large language models},
  author={Liu, Aiwei and Pan, Leyi and Hu, Xuming and Li, Shuang and Wen, Lijie and King, Irwin and Philip, S Yu},
  booktitle={ICLR},
  year={2024}
}

@inproceedings{
hu2024unbiased,
title={Unbiased Watermark for Large Language Models},
author={Zhengmian Hu and Lichang Chen and Xidong Wu and Yihan Wu and Hongyang Zhang and Heng Huang},
booktitle={ICLR},
year={2024}
}

@inproceedings{
yoo2024advancing,
title={Advancing Beyond Identification: Multi-bit Watermark for Large Language Models},
author={KiYoon Yoo and Wonhyuk Ahn and Nojun Kwak},
year={2024},
booktitle={NAACL}
}

@inproceedings{
zhao2024provable,
title={Provable Robust Watermarking for {AI}-Generated Text},
author={Xuandong Zhao and Prabhanjan Vijendra Ananth and Lei Li and Yu-Xiang Wang},
booktitle={ICLR},
year={2024}
}

@inproceedings{jovanovic2024watermark,
author = {Jovanovi\'{c}, Nikola and Staab, Robin and Vechev, Martin},
title = {Watermark stealing in large language models},
year = {2024},
booktitle = {ICML},
articleno = {907},
numpages = {24},
}

@inproceedings{wu2024bypassing,
  title={Bypassing {LLM WMs} with color-aware substitutions},
  author={Wu, Qilong and Chandrasekaran, Varun},
  booktitle={ACL},
  year={2024}
}

@article{liu2024survey,
  title={A survey of text watermarking in the era of large language models},
  author={Liu, Aiwei and Pan, Leyi and Lu, Yijian and Li, Jingjing and Hu, Xuming and Zhang, Xi and Wen, Lijie and King, Irwin and Xiong, Hui and Yu, Philip},
  journal={ACM Computing Surveys},
  year={2024},
  publisher={ACM New York, NY}
}

@article{zhang2024emmark,
  title={EmMark: Robust Watermarks for IP Protection of Embedded Quantized Large Language Models},
  author={Zhang, Ruisi and Koushanfar, Farinaz},
  journal={arXiv},
  year={2024}
}

@inproceedings{baldassini2024cross,
  title={Cross-Attention watermarking of Large Language Models},
  author={Baldassini, Folco Bertini and Nguyen, Huy H and Chang, Ching-Chung and Echizen, Isao},
  booktitle={ICASSP},
  year={2024},
}

@inproceedings{molenda2024waterjudge,
  title={Waterjudge: Quality-detection trade-off when watermarking large language models},
  author={Molenda, Piotr and Liusie, Adian and Gales, Mark JF},
  booktitle={NAACL},
  year={2024}
}

@article{fu2025multi,
  title={Multi-use llm watermarking and the false detection problem},
  author={Fu, Zihao and Russell, Chris},
  journal={arXiv},
  year={2025}
}

@article{luan2026vow,
  title={VOW: Verifiable and Oblivious Watermark Detection for Large Language Models},
  author={Luan, Xiaokun and Zhang, Yihao and Su, Pengcheng and Lei, Feiran and Sun, Meng},
  journal={arXiv},
  year={2026}
}

@article{aremu2026watermarking,
  title={Watermarking Should Be Treated as a Monitoring Primitive},
  author={Aremu, Toluwani and Lukas, Nils and Zhang, Jie},
  journal={arXiv},
  year={2026}
}

@inproceedings{dabiriaghdam2025simmark,
    title = "{S}im{M}ark: A Robust Sentence-Level Similarity-Based Watermarking Algorithm for Large Language Models",
    author = "Dabiriaghdam, Amirhossein  and
      Wang, Lele",
    booktitle = "EMNLP",
    year = "2025",
    publisher = "Association for Computational Linguistics",
}

@inproceedings{kwon2023efficient,
  title={Efficient Memory Management for{LLM} Serving with PagedAttention},
    author={Kwon, Woosuk and Li, Zhuohan and Zhuang, Siyuan and Sheng, Ying and Zheng, Lianmin and Yu, Cody Hao and Gonzalez, Joseph and Zhang, Hao and Stoica, Ion},
  booktitle={ACM SIGOPS},
  year={2023}
}

@article{agrawal2023sarathi,
  title={Sarathi: Efficient llm inference by piggybacking decodes with chunked prefills},
  author={Agrawal, Amey and Panwar, Ashish and Mohan, Jayashree and Kwatra, Nipun and Gulavani, Bhargav S and Ramjee, Ramachandran},
  journal={arXiv},
  year={2023}
}

@inproceedings{zhong2024distserve,
  title={ {DistServe} : Disaggregating prefill and decoding for goodput-optimized {LLM} serving},
  author={Zhong, Yinmin and Liu, Shengyu and Chen, Junda and Hu, Jianbo and Zhu, Yibo and Liu, Xuanzhe and Jin, Xin and Zhang, Hao},
  booktitle={OSDI},
  year={2024}
}

@inproceedings{pan2025waterseeker,
  title={Waterseeker: Pioneering efficient detection of watermarked segments in large documents},
  author={Pan, Leyi and Liu, Aiwei and Lu, Yijian and Gao, Zitian and Di, Yichen and Wen, Lijie and King, Irwin and Yu, Philip S},
  booktitle={Findings of the Association for Computational Linguistics: NAACL 2025},
  pages={2866--2882},
  year={2025}
}

@article{akidau2015dataflow,
  title={The dataflow model: a practical approach to balancing correctness, latency, and cost in massive-scale, unbounded, out-of-order data processing},
  author={Akidau, Tyler and Bradshaw, Robert and Chambers, Craig and Chernyak, Slava and Fern{\'a}ndez-Moctezuma, Rafael J and Lax, Reuven and McVeety, Sam and Mills, Daniel and Perry, Frances and Schmidt, Eric and others},
  journal={Proceedings of the VLDB Endowment},
  year={2015},
  publisher={VLDB Endowment}
}

@inproceedings{zhao2025efficiently,
  title={Efficiently identifying watermarked segments in mixed-source texts},
  author={Zhao, Xuandong and Liao, Chenwen and Wang, Yu-Xiang and Li, Lei},
  booktitle={ACL},
  year={2025}
}

@article{zhang2026character,
  title={Character-level perturbations disrupt llm watermarks},
  author={Zhang, Zhaoxi and Zhang, Xiaomei and Zhang, Yanjun and Zhang, He and Pan, Shirui and Liu, Bo and others},
  journal={NDSS},
  year={2026}
}

@article{aremu2026mitigating,
      title={Mitigating Watermark Forgery in Generative Models via Randomized Key Selection}, 
      author={Toluwani Aremu and Noor Hussein and Munachiso Nwadike and Samuele Poppi and Jie Zhang and Karthik Nandakumar and Neil Gong and Nils Lukas},
      year={2026},
      journal={arXiv}
}

@inproceedings{pan2025canllm,
    title = "Can {LLM} Watermarks Robustly Prevent Unauthorized Knowledge Distillation?",
    author = "Pan, Leyi  and
      Liu, Aiwei  and
      Huang, Shiyu  and
      Lu, Yijian  and
      Hu, Xuming  and
      Wen, Lijie  and
      King, Irwin  and
      Yu, Philip S.",
    booktitle = "ACL",
    year = "2025",
}

@article{shao2025explanation,
  title={Explanation as a watermark: Towards harmless and multi-bit model ownership verification via watermarking feature attribution},
  author={Shao, Shuo and Li, Yiming and Yao, Hongwei and He, Yiling and Qin, Zhan and Ren, Kui},
  journal={NDSS},
  year={2025}
}

@article{yang2023watermarking,
  title={Watermarking text generated by black-box language models},
  author={Yang, Xi and Chen, Kejiang and Zhang, Weiming and Liu, Chang and Qi, Yuang and Zhang, Jie and Fang, Han and Yu, Nenghai},
  journal={arXiv},
  year={2023}
}

@article{meral2009natural,
  title={Natural language watermarking via morphosyntactic alterations},
  author={Meral, Hasan Mesut and Sankur, B{\"u}lent and {\"O}zsoy, A Sumru and G{\"u}ng{\"o}r, Tunga and Sevin{\c{c}}, Emre},
  journal={Computer Speech \& Language},
  volume={23},
  number={1},
  pages={107--125},
  year={2009},
  publisher={Elsevier}
}

@article{ahvanooey2018aitsteg,
  title={AITSteg: An innovative text steganography technique for hidden transmission of text message via social media},
  author={Ahvanooey, Milad Taleby and Li, Qianmu and Hou, Jun and Mazraeh, Hassan Dana and Zhang, Jing},
  journal={IEEE Access},
  year={2018},
  publisher={IEEE}
}

@article{sato2023embarrassingly,
  title={Embarrassingly simple text watermarks},
  author={Sato, Ryoma and Takezawa, Yuki and Bao, Han and Niwa, Kenta and Yamada, Makoto},
  journal={arXiv},
  year={2023}
}

@article{wang2024building,
  title={Building intelligence identification system via large language model watermarking: a survey and beyond},
  author={Wang, Xuhong and Jiang, Haoyu and Yu, Yi and Yu, Jingru and Lin, Yilun and Yi, Ping and Wang, Yingchun and Qiao, Yu and Li, Li and Wang, Fei-Yue},
  journal={Artificial Intelligence Review},
  volume={58},
  number={8},
  pages={249},
  year={2025},
  publisher={Springer}
}

@article{pang2025modelshield,
  title={ModelShield: Adaptive and Robust Watermark against Model Extraction Attack},
  author={Pang, Kaiyi and Qi, Tao and Wu, Chuhan and Bai, Minhao and Jiang, Minghu and Huang, Yongfeng},
  journal={TIFS},
  year={2025},
}

@article{pang2024no,
  title={No free lunch in llm watermarking: Trade-offs in watermarking design choices},
  author={Pang, Qi and Hu, Shengyuan and Zheng, Wenting and Smith, Virginia},
  journal={NeurIPS},
  year={2024}
}

@article{
he2022cater,
  title={Cater: {IP} protection on text generation apis via conditional watermarks},
  author={He, Xuanli and Xu, Qiongkai and Zeng, Yi and Lyu, Lingjuan and others},
  journal={Advances in Neural Information Processing Systems},
  year={2022}
}

@inproceedings{chang2024postmark,
    title = "{P}ost{M}ark: A Robust Blackbox Watermark for Large Language Models",
    author = "Chang, Yapei  and
      Krishna, Kalpesh  and
      Houmansadr, Amir  and
      Wieting, John Frederick  and
      Iyyer, Mohit",
    booktitle = "EMNLP",
    year = "2024"
}

@article{duan2025pvmark,
  title={Pvmark: Enabling public verifiability for llm watermarking schemes},
  author={Duan, Haohua and Xiang, Liyao and Zhang, Xin},
  journal={USENIX Security},
  year={2026}
}

@inproceedings{hou2024semstamp,
  title={Semstamp: A semantic watermark with paraphrastic robustness for text generation},
      author = "Hou, Abe  and
      Zhang, Jingyu  and
      He, Tianxing  and
      Wang, Yichen  and
      Chuang, Yung-Sung  and
      Wang, Hongwei  and
      Shen, Lingfeng  and
      Van Durme, Benjamin  and
      others",
  booktitle={NAACL},
  year={2024}
}

@inproceedings{kirchenbauer2023reliability,
  title={On the reliability of watermarks for large language models},
  author={Kirchenbauer, John and Geiping, Jonas and Wen, Yuxin and Shu, Manli and Saifullah, Khalid and Kong, Kezhi and   others},
  booktitle={ACL},
  year={2024}
}

@inproceedings{pasquini2025llmmap,
  title={LLMmap: Fingerprinting for large language models},
  author={Pasquini, Dario and Kornaropoulos, Evgenios M and Ateniese, Giuseppe},
  booktitle={USENIX Security},
  year={2025}
}

@inproceedings{dang2025delta,
  title={$\delta$-STEAL: LLM Stealing Attack with Local Differential Privacy},
  author={Dang, Kieu and Lai, Phung and Phan, NhatHai and Shen, Yelong and Jin, Ruoming and Khreishah, Abdallah},
  booktitle={ACML},
  year={2025}
}

@inproceedings{FengXiaoyan2025BUMW,
author = {Feng, Xiaoyan and Zhang, He and Zhang, Yanjun and Zhang, Leo Yu and Pan, Shirui},
title = {BiMark: Unbiased Multilayer Watermarking for Large Language Models},
booktitle ={ICML},
year = {2025}
}

@article{LiZhuang2025BETW,
  title={BiMarker: Enhancing text watermark detection for large language models with bipolar watermarks},
  author={Li, Zhuang and Yi, Qiuping and Ji, Zongcheng and Lu, Yijian and Zou, Shun and Li, Yanqi and Xiao, Keyang and Liang, Hongliang},
  journal={Neurocomputing},
  pages={133843},
  year={2026},
  publisher={Elsevier}
}

@article{ZhangJunyan2025CANS,
author = {Zhang, Junyan and Liu, Shuliang and Liu, Aiwei and Gao, Yubo and Li, Jungang and Gu, Xiaojie and Hu, Xuming},
copyright = {http://arxiv.org/licenses/nonexclusive-distrib/1.0},
language = {eng},
title = {CoheMark: A Novel Sentence-Level Watermark for Enhanced Text Quality},
journal= {ICLR},
year = {2025},
}

@article{xue2025pro,
  title={PRO: Enabling Precise and Robust Text Watermark for Open-Source LLMs},
  author={Xue, Jiaqi and Zhao, Yifei and Ghanim, Mansour Al and Gao, Shangqian and Sun, Ruimin and Lou, Qian and Zheng, Mengxin},
  journal={NeurIPS},
  year={2026}
}

@inproceedings{hellmeier2025hidden,
  title={A hidden digital text watermarking method using unicode whitespace replacement},
  author={Hellmeier, Malte and Qarawlus, Haydar and Norkowski, Hendrik and Howar, Falk},
booktitle={Hawaii International Conference on System Sciences},
  year={2025}
}

@article{LiHao2025FEtD,
  title={From Essence to Defense: Adaptive Semantic-aware Watermarking for Embedding-as-a-Service Copyright Protection},
  author={Li, Hao and Ren, Yubing and Cao, Yanan and Li, Yingjie and Fang, Fang and Wang, Xuebin},
  journal={arXiv preprint arXiv:2512.16439},
  year={2025}
}

@article{hellmeier2025innamark,
  author={Hellmeier, Malte and Norkowski, Hendrik and Schrewe, Ernst-Christoph and Qarawlus, Haydar and Howar, Falk},
  journal={IEEE Access}, 
  title={Innamark: A Whitespace Replacement Information-Hiding Method}, 
  year={2025},
  volume={13},
}

@inproceedings{li2024identifying,
  title={Where am i from? identifying origin of llm-generated content},
  author={Li, Liying and Bai, Yihan and Cheng, Minhao},
  booktitle={EMNLP},
  pages={12218--12229},
  year={2024}
}

@inproceedings{gu2025invisible,
  title={Invisible Entropy: Towards Safe and Efficient Low-Entropy LLM Watermarking},
author = "Gu, Tianle  and
      Wang, Zongqi  and
      Huang, Kexin  and
      Yao, Yuanqi  and
      Zhang, Xiangliang  and
      Yang, Yujiu  and
      Chen, Xiuying",
    booktitle = "Proceedings of EMNLP",
    year = "2025"
}

@inproceedings{cohen2025watermarking,
  title={Watermarking language models for many adaptive users},
  author={Cohen, Aloni and Hoover, Alexander and Schoenbach, Gabe},
  booktitle={2025 IEEE Symposium on Security and Privacy (SP)},
  pages={2583--2601},
  year={2025},
  organization={IEEE}
}

@article{xu2024robust,
  title={Robust Multi-bit Text Watermark with LLM-based Paraphrasers},
  author={Xu, Xiaojun and Jia, Jinghan and Yao, Yuanshun and Liu, Yang and Li, Hang},
  journal={arXiv preprint arXiv:2412.03123},
  year={2024}
}

@inproceedings{liu2026incentives,
  title={Position: LLM Watermarking Should Align Stakeholders' Incentives for Practical Adoption},
  author={Liu, Yepeng and Zhao, Xuandong and Song, Dawn and Wornell, Gregory W. and Bu, Yuheng},
  booktitle={ACL},
  year={2026},
  publisher={Association for Computational Linguistics},
}

@article{zhou2025survey,
  title={A Survey of LLM x DATA},
  author={Zhou, Xuanhe and He, Junxuan and Zhou, Wei and Chen, Haodong and Tang, Zirui and Zhao, Haoyu and Tong, Xin and Li, Guoliang and Chen, Youmin and Zhou, Jun and others},
  journal={arXiv},
  year={2025}
}

@article{fernandez2023large,
  title={How large language models will disrupt data management},
  author={Fernandez, Raul Castro and Elmore, Aaron J and Franklin, Michael J and Krishnan, Sanjay and Tan, Chenhao},
  journal={Proceedings of the VLDB Endowment},
  volume={16},
  number={11},
  pages={3302--3309},
  year={2023},
  publisher={VLDB Endowment}
}

@inproceedings{dai2024neural,
  title={Neural retrievers are biased towards llm-generated content},
  author={Dai, Sunhao and Zhou, Yuqi and Pang, Liang and Liu, Weihao and Hu, Xiaolin and Liu, Yong and Zhang, Xiao and Wang, Gang and Xu, Jun},
  booktitle={ACM SIGKDD},
  pages={526--537},
  year={2024}
}

@article{shumailov2024ai,
  title={AI models collapse when trained on recursively generated data},
  author={Shumailov, Ilia and Shumaylov, Zakhar and Zhao, Yiren and Papernot, Nicolas and Anderson, Ross and Gal, Yarin},
  journal={Nature},
  volume={631},
  number={8022},
  pages={755--759},
  year={2024},
  publisher={Nature Publishing Group UK London}
}

@article{jaiswal2025serving,
  title={Serving models fast and slow: optimizing heterogeneous llm inferencing workloads at scale},
  author={Jaiswal, Shashwat and Jain, Kunal and Simmhan, Yogesh and Parayil, Anjaly and Mallick, Ankur and Wang, Rujia and Amant, Renee St and Bansal, Chetan and R{\"u}hle, Victor and Kulkarni, Anoop and others},
  journal={ACM Sigmetrics},
  year={2026}
}

@article{cai2025optimal,
  title={Optimal detection for language watermarks with pseudorandom collision},
  author={Cai, T Tony and Li, Xiang and Long, Qi and Su, Weijie J and Wen, Garrett G},
  journal={arXiv},
  year={2025}
}

@inproceedings{cheng2025revealing,
  title={Revealing weaknesses in text watermarking through self-information rewrite attacks},
  author={Cheng, Yixin and Guo, Hongcheng and Li, Yangming and Sigal, Leonid},
  booktitle={ICML},
  year={2025}
}

\end{document}